\documentstyle[twoside,fleqn,espcrc2,epsfig]{article}
\def\ppb{p\bar{p}}
\def\as{\alpha_s}
\def\gl{\tilde{g}}
\def\sq{\tilde{q}}
\def\sqb{\bar{\tilde{q}}}
\def\qb{\bar{q}}
\def\ms{m_{\tilde q}}
\def\mg{m_{\tilde g}}
\def\eps{\varepsilon}
\def\DR{$\overline{DR}$~}
\def\MS{$\overline{MS}$~}
\def\ghat{\hat{g}_s}
\def\lsim{\;\raisebox{-.3em}{$\stackrel{\displaystyle <}{\sim}$}\;}

\newcommand{\zp}[3]{{\sl Z. Phys.} {\bf #1} (19#2) #3}

\newcommand{\pl}[3]{{\sl Phys. Lett.} {\bf #1} (19#2) #3}
\newcommand{\pr}[3]{{\sl Phys. Rev.} {\bf #1} (19#2) #3}
\newcommand{\prl}[3]{{\sl Phys. Rev. Lett.} {\bf #1} (19#2) #3}
\newcommand{\hep}[1]{{\sl hep--ph/}{#1}}
\newcommand{\desy}[1]{{\sl DESY-Report~}{#1}}
\title{SUSY-QCD corrections in the squark--gluino sector}
\author{W. Beenakker%
        \address{Instituut--Lorentz, University of Leiden,
                 P.O.~Box 9506, NL--2300 RA Leiden, The Netherlands.}%
        \thanks{Partially supported by EU contract CHRX-CT-92-0004.}%
        \thanks{Research supported by a fellowship of the Royal Dutch Academy 
                of Arts and Sciences.}
        and 
        R. H\"opker%
        \address{DESY, Theory Division, Notkestrasse 85,
                 D--22603 Hamburg, Germany.}
       }
\begin{document}
\begin{abstract}
A status report is given of the calculations of next-to-leading-order ($N=1$)
supersymmetric QCD corrections to the production of squarks and gluinos in
$\ppb/pp$ collisions. The implementation of these SUSY-QCD corrections 
leads to more stable theoretical predictions and to a substantial increase of
the production cross-sections. In addition we give a discussion of the use of
the \MS scheme for renormalizing the coupling constants in the QCD sector of
($N=1$) supersymmetric theories.
\end{abstract}
%
\maketitle
\section{INTRODUCTION}

The colored squarks ($\sq_L,\sq_R$) and gluinos ($\gl$), the supersymmetric 
partners of the quarks ($q$) and gluons ($g$), can be searched for most 
efficiently at high-energy hadron colliders. As R-parity is conserved in the 
QCD sector of ($N=1$) supersymmetric theories, these particles are always 
produced in pairs. At the moment they can be searched for at the Fermilab 
Tevatron, a $\ppb$ collider with a centre-of-mass energy of 1.8~TeV. In the 
future the CERN Large Hadron Collider (LHC), a $pp$ collider with an envisaged 
centre-of-mass energy of 14~TeV, will allow to cover mass values up to 
1--2~TeV. 

\begin{figure}[tb]
  \begin{center}
    \vspace*{-0.2cm}  
    \hspace*{-0.1cm}
    \epsfig{file=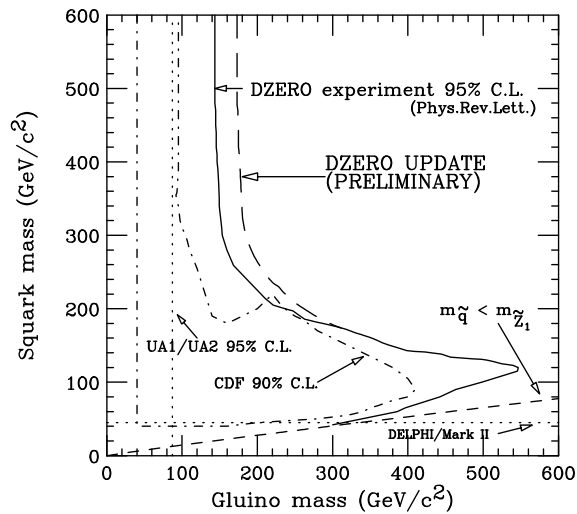,width=7.5cm}
    \vspace*{-1.3cm}  
  \end{center}
  \caption[]{Experimental lower mass bounds for squarks ($\sq\neq\tilde{t}$)
             and gluinos from the Tevatron~\cite{d0}.}
  \label{fig:masslimit1}
\end{figure}

At present the main search strategy employed at the Tevatron involves the 
search for jets and missing transverse energy \cite{cdf1,d0}. R-parity 
conserving decay cascades of the squarks and gluinos result in Standard-Model 
hadrons, i.e., jets, and LSP's ($\chi_1^0$), responsible for the missing 
energy. An example of a complete process, involving production and subsequent 
decay, is given by 
\begin{displaymath}
    \ppb \rightarrow \sq_L+\sqb_R \rightarrow
    \chi^+ q' + \chi^0_1 \qb \rightarrow
    \chi^0_1 q \qb' q' + \chi^0_1 \qb, 
\end{displaymath}
with $q'$ the isospin partner of $q$.
The present status of the squark and gluino searches, based on this strategy, 
is depicted in Fig.~\ref{fig:masslimit1}. As the search has been negative up 
to now, only lower bounds on the squark and gluino masses are given.
The present lower mass bounds, derived from Fig.~\ref{fig:masslimit1}, are 
($\sq\neq\tilde{t}$)
\begin{eqnarray*}
  \mg & \ge & 175~\mbox{GeV}, \\
  \ms & \ge & 175~\mbox{GeV} \quad \mbox{for} \quad \mg \le 300~\mbox{GeV}.
\end{eqnarray*}

So far this analysis has been based on the lowest-order (LO) production 
cross-sections. For obtaining adequate theoretical predictions the LO 
cross-sections are in general not sufficient. The most important arguments in 
favor of an analysis that takes into account the next-to-leading-order (NLO) 
SUSY-QCD corrections are:
\begin{itemize}
  \item The LO cross-sections have a strong dependence on the
        {\it a priori} unknown renormalization scale $Q_R$.
        Consequently, the theoretical predictions have 
        in general an uncertainty that is almost as large as the cross-section
        itself. By implementing the NLO corrections a 
        substantial reduction of the scale dependence is expected.
  \item From experience with similar processes (e.g.~hadroproduction of top
        quarks), the NLO QCD corrections are expected to be of the order of
        +30\%. 
  \item An enhancement of the cross-section would lead to a higher value for 
        the lower mass bounds for squarks and gluinos.
  \item In case of discovery of squarks and gluinos, a precise knowledge of the
        total cross-sections is required for the determination of the masses
        of the particles. In contrast to the production of e.g.~top-quark or
        Z-boson pairs, it seems unlikely that the masses of squarks and 
        gluinos can be determined by means of reconstruction (in view of the 
        invisible LSP's). Probably they can only be inferred from the 
        experimental production rates and the theoretical prediction for the 
        total cross-section.
\end{itemize}

Here we report on the status of the calculations of NLO SUSY-QCD corrections 
to the production of (on-shell) squarks and gluinos in $p\bar{p}/pp$
collisions, based on the studies presented in \cite{BHSZprod}.
For a more comprehensive report on this topic we refer to \cite{BHSZreport}.

\section{TECHNICAL SET-UP}

\begin{figure}[htb]
  \begin{center}
  \vspace*{-1.4cm}
  \hspace*{-2.4cm}
  \epsfig{file=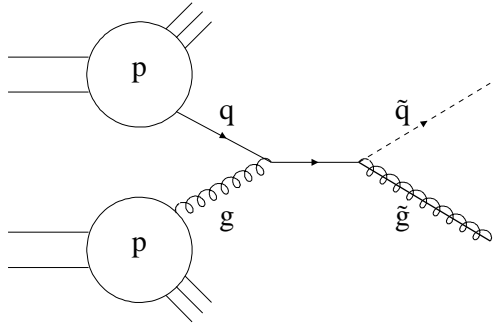,width=12cm}
  \vspace*{-3.0cm}
  \end{center}
  \caption[]{Generic Feynman diagram for the production of squarks and gluinos 
             in $\ppb/pp$ collisions.}
  \label{fig:hadronprod}
\end{figure}
We consider the following hadronic production processes (generically depicted
in Fig.~\ref{fig:hadronprod}):
\begin{equation}
 \ppb/pp \to \sq\sqb,\,\sq\sq,\,\gl\gl,\,\sq\gl \qquad (\sq\neq\tilde{t}),
\end{equation}
where the chiralities and flavors of the squarks 
(e.g.~$\tilde{u}_L,\,\tilde{d}_R$) as well as the charge-conjugate
final states (e.g.~$\sqb\sqb$) are implicitly summed over. In analogy to the 
experimental analysis, we exclude top-squarks from the final state and take 
all produced squarks to be mass degenerate. A study of the production of pairs 
of top-squarks is in progress, including the (model-dependent) mixing effects 
in the squark sector. At the partonic level (right-hand side of 
Fig.~\ref{fig:hadronprod}) many different subprocesses contribute at LO and 
NLO, corresponding to different flavors/chiralities of the squarks and 
different initial-state partons. The initial-state partons are made up of the 
massless gluons and the five light quark flavors ($n_f=5$), which are 
considered to be massless as well. Note that not all initial states are
possible for a given final state. At LO, for instance, the production of 
squark--antisquark final states requires quark--antiquark or gluon--gluon 
initial states, whereas the squark--gluino final states are only possible in 
quark--gluon reactions.

The NLO SUSY-QCD corrections comprise the virtual corrections (consisting of 
self-energy corrections, vertex corrections, and box diagrams), 
real-gluon radiation (with an additional gluon attached to the LO diagrams),
and the radiation of a massless quark (opening additional initial-state
channels: e.g.~$g q \to \sq \sqb q$).

\begin{figure*}[tb]
  \begin{center}
  \vspace*{-1.8cm}
  \hspace*{-0cm}
  \epsfig{file=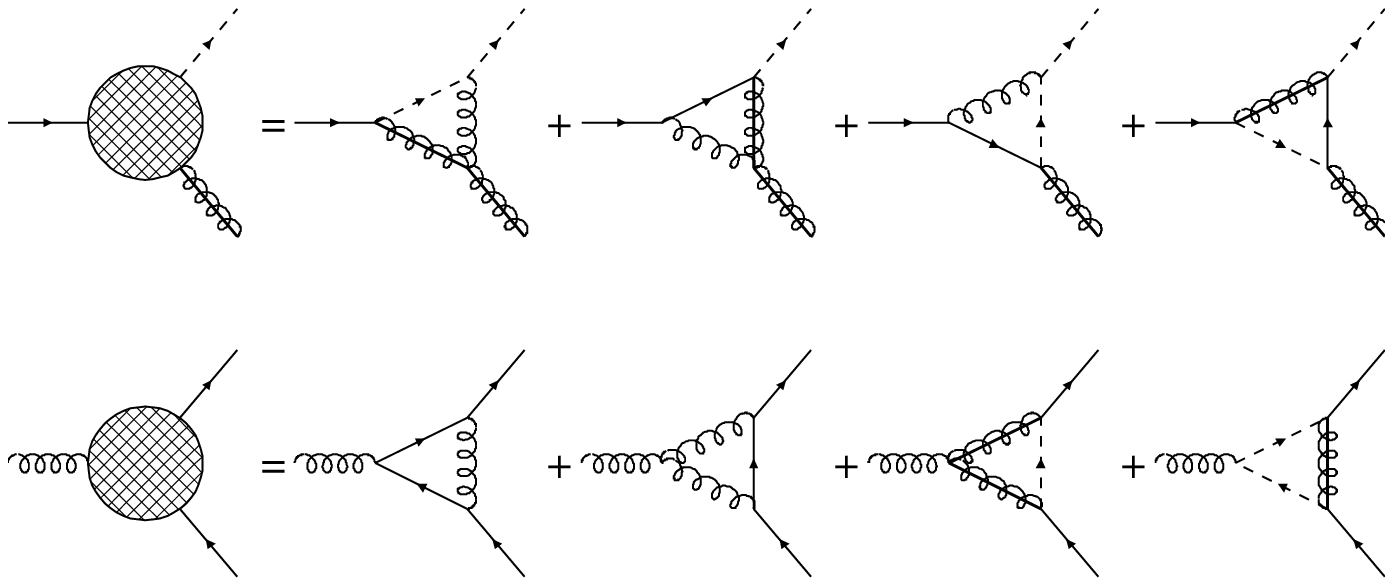,width=15cm}
  \vspace*{-2.5cm}
  \end{center}
  \begin{picture}(0,0)
    \setlength{\unitlength}{1cm}
    \put(1.7,1.0){\makebox(0,0){$g$}}
    \put(3.7,2.0){\makebox(0,0){$q$}}
    \put(3.7,0.0){\makebox(0,0){$q$}}
    \put(1.7,3.9){\makebox(0,0){$q$}}
    \put(3.7,4.9){\makebox(0,0){$\sq$}}
    \put(3.7,2.9){\makebox(0,0){$\gl$}}
  \end{picture}
  \vspace*{-0.5cm}
  \caption[]{Feynman diagrams for the virtual NLO SUSY-QCD corrections to the
             quark--squark--gluino vertex (Yukawa coupling) and the 
             quark--quark--gluon vertex (gauge coupling).}
  \label{fig:feynvirt}
\end{figure*}
For the particles inside the loops we use the complete supersymmetric QCD
spectrum, i.e., gluons, gluinos, all quarks, and all squarks. We have 
excluded the top-squarks from the final states. In order to have a 
consistent NLO calculation, however, we have to take these top-squarks into
account inside loops. For the sake of simplicity we take them to be 
mass-degenerate with the other squarks. For the top quark we use the mass 
$m_t=175$~GeV. Consequently the final results will depend on two free 
parameters: the squark mass $\ms$ and the gluino mass $\mg$.

For the internal gluon propagators we use the Feynman gauge. As a result,
Faddeev--Popov ghost contributions have to be taken into account in the 
gluon self-energy and the three-gluon vertex corrections. For the external
gluon lines only the transverse polarizations are needed. 

The divergences appearing in the NLO corrections are regularized by 
performing the calculations in $n=4-2\eps$ dimensions. These divergences
consist of ultraviolet (UV), infra-red (IR), and collinear divergences, and 
show up as poles of the form $\eps^{-i}$~($i=1,2$). For the treatment of the 
$\gamma_5$ Dirac matrix, entering through the quark--squark--gluino Yukawa 
couplings, we use the `naive' scheme. This involves a $\gamma_5$ that 
anticommutes with the other gamma matrices. This is a legitimate scheme at 
the one-loop level for anomaly-free theories.

The UV divergences can be removed by renormalizing the coupling constants and
the masses of the massive particles. In the case of the mass renormalization 
we have opted for an on-shell scheme with real subtraction point, involving 
the subtraction of the real part of the on-shell self-energies at the 
real-valued pole masses. For the renormalization of the QCD coupling constant  
one usually resorts to the modified Minimal Subtraction (\MS) scheme. The \MS 
scheme involves $n$-dimensional regularization, i.e., treating fields, phase 
space, and loop momenta in $n$ dimensions. The UV $1/\eps$ poles are 
subtracted, together with specific transcendental constants, 
at an {\it a priori} 
arbitrary subtraction point (renormalization scale) $Q_R$ . In supersymmetric 
theories, however, a complication occurs. In $n\neq 4$ dimensions the \MS 
scheme introduces a mismatch between the number of gluon ($n-2$) and gluino (2)
degrees of freedom. As these ${\cal O}(\eps)$ mismatches will result in finite 
contributions, the \MS scheme violates supersymmetry explicitly. In particular,
the $q\sq\gl$ Yukawa coupling $\ghat$, which by supersymmetry should coincide 
with the $qqg$ gauge coupling $g_s$, deviates from $g_s$ by a finite amount at 
the one-loop order. 
Requiring the physical amplitudes to be independent of the renormalization 
scheme, a shift between the bare Yukawa and gauge couplings must be introduced
in the \MS scheme, 
\begin{small}
\begin{equation}
  \hat{g}_s = g_s \left[ 1 + \frac{\as}{4\pi}\left(\frac{2}{3}N
              -\frac{1}{2}C_F\right)\right] 
            = g_s \left[ 1 + \frac{\as}{3\pi}\right],
  \label{finshift}
\end{equation}
\end{small}
which effectively subtracts the contributions of the false, non-supersymmetric
degrees of freedom (also called $\eps$ scalars) \cite{martin}. Here we used 
$\as=g_s^2/(4\pi),\,N=3$, and $C_F=(N^2-1)/(2N)=4/3$.

The need for introducing a finite shift is best demonstrated for the effective
(one-loop corrected) Yukawa coupling (see Fig.~\ref{fig:feynvirt}), which 
must be equal to the effective gauge coupling in an exact supersymmetric world 
with massless gluons/gluinos and equal-mass quarks/squarks. For the sake of 
simplicity we define the one-loop corrected effective couplings 
$\Gamma_s^{\mbox{\scriptsize eff}}(Q^2) = g_s\,[1+\delta(Q^2)]$ and 
$\hat{\Gamma}_s^{\mbox{\scriptsize eff}}(Q^2) = \ghat\,[1+\hat{\delta}(Q^2)]$ 
in the limit of on-shell quarks/squarks and almost on-shell gluons/gluinos, 
with virtuality $Q^2\ll \ms^2=m_q^2$. In this limit the couplings do not 
contain gauge-dependent terms. In the \MS scheme we find after charge 
renormalization \cite{BHZdecay}
\begin{small}
\begin{equation}
  \overline{MS}: \ 
  \frac{\hat{\Gamma}_s^{\mbox{\scriptsize eff}}(Q^2)}{\ghat} = 
  \frac{\Gamma_s^{\mbox{\scriptsize eff}}(Q^2)}{g_s} 
  + \frac{\as}{4\pi} \left( \frac{1}{2}C_F - \frac{2}{3}N \right). 
  \label{qsqgltot}
\end{equation}
\end{small}
The difference between the two effective couplings coincides with the 
shift~(\ref{finshift}). Taking into account this finite shift of the bare
couplings in the \MS scheme, both effective couplings become identical at the 
one-loop level. In this way supersymmetry is preserved and the \MS scheme
becomes a workable one.

An alternative renormalization~scheme is the modified Dimensional 
Reduction (\DR) scheme. This scheme consists in treating the fields in 4 
dimensions and the phase space and loop momenta in $n$ dimensions. As such no 
mismatch is introduced and supersymmetry is preserved. At the level of the 
above-defined effective couplings this is reflected in the equality 
\cite{BHZdecay}
\begin{small}
\begin{equation}
  \overline{DR}: \ \ 
  \frac{\hat{\Gamma}_s^{\mbox{\scriptsize eff}}(Q^2)}{\ghat} = 
  \frac{\Gamma_s^{\mbox{\scriptsize eff}}(Q^2)}{g_s}.
\end{equation}
\end{small}
As a result, both couplings are identical order by order.
It should be noted that the transition from the effective gauge coupling in
\MS to the one in \DR involves a well-known finite renormalization 
$(\as N)/(24\pi)=\as/(8\pi)$.

In the following we use the \MS renormalization scheme in combination with the 
finite shift of the Yukawa coupling. In this way supersymmetry is preserved
on the one hand, while on the other hand the definition of the strong gauge
coupling corresponds to the usual Standard-Model measurements. In addition to 
the poles also some logarithms are subtracted in order to decouple the massive 
particles (top quarks, squarks, gluinos) from the running of 
$\as(Q_R^2)$. The $Q_R^2$ evolution of the strong coupling is in this
decoupling renormalization scheme completely determined by the light-particle
spectrum (gluons and $n_f=5$ massless quarks): 
\begin{small} 
\begin{equation}
  \frac{\partial\, g_s^2(Q_R^2)}{\partial\log(Q_R^2)}
  = -\as^2(Q_R^2)\,\left[ \frac{11}{3}\,N -\frac{2}{3}\,n_f \right].
\end{equation}
\end{small}

The above described methods to renormalize the UV divergences result in 
cross-sections that are UV finite. Nevertheless there are still divergences
left. The IR ones will cancel in the sum of virtual corrections and soft-gluon
radiation. In order to separate soft from hard radiation a cut-off $\Delta$ is 
introduced in the invariant mass corresponding to the radiated gluon and one 
of the produced massive particles. If soft and hard  
contributions are added up, any $\Delta$ dependence disappears from the 
cross-sections for $\Delta \to 0$. The remaining collinear singularities, 
finally, can be absorbed into the renormalization of the parton densities, 
carried out in the \MS mass-factorization scheme. This introduces yet another
{\it a priori} unknown scale, the factorization scale $Q_F$.

In the context of mass factorization an interesting observation was made 
during the comparison of the \MS and \DR results. Even after correcting for 
$\eps$-scalar contributions, leading to finite differences 
between the splitting functions in \MS and \DR \cite{korner}, discrepancies of 
${\cal O}(\ms^2,\mg^2)$ persisted. The exact source of these discrepancies is 
under investigation. This phenomenon is, however, not an artefact of 
supersymmetry, since it is also observed in the (pure QCD) process of 
top-quark production \cite{heavyfl1}.

\begin{figure}[hbt]
  \begin{center}
    \vspace*{-1.0cm}
    \hspace*{-2.0cm}
    \epsfig{file=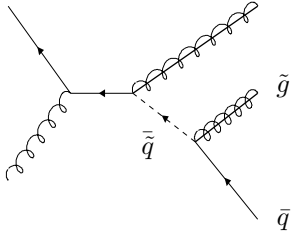,width=12cm}
    \vspace*{-1.5cm}
  \end{center}
  \begin{picture}(0,0)
    \setlength{\unitlength}{1cm}
    \put(3.6,1.0){\makebox(0,0){$\sqb$}}
    \put(5.4,1.8){\makebox(0,0){$\gl$}}
    \put(5.4,0.0){\makebox(0,0){$\qb$}}
  \end{picture}
  \vspace*{-0.5cm}
  \caption[]{Example of a NLO Feynman diagram that can give rise to an on-shell
             intermediate squark state.}
  \label{fig:subtract}
\end{figure}

The last singularity we have to deal with is of kinematical nature.
If the gluinos are lighter than the squarks, on-shell squarks can decay into 
massless quarks plus gluinos ($\sq \to q\gl $). This means that inside the 
phase space of massless-quark-radiation processes like $g\qb \to \gl\gl\qb$ 
(see Fig.~\ref{fig:subtract}) explicit particle poles can emerge. 
These particle poles, however, correspond to on-shell (LO) production of the 
intermediate on-shell squark state, with subsequent (LO) decay into a massless 
quark plus gluino. In our narrow-width approach, consisting in neglecting the 
finite decay widths of the squarks/gluinos whenever possible, these situations 
are already accounted for by the LO cross-sections (e.g. squark--gluino 
production in Fig.~\ref{fig:subtract}).
In order to avoid double counting these kinematical situations have to be 
subtracted from the NLO corrections. Similar subtractions have to be performed 
if the squarks are lighter than the gluinos. For more details concerning the
exact subtraction procedure we refer to \cite{BHSZreport}. 

\section{RESULTS}

\begin{figure}[b]
  \begin{center}
    \vspace{-1.2cm}
    \hspace*{-1.4cm}
    \epsfig{file=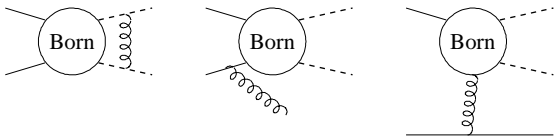,width=10cm}
    \vspace{-1.9cm}
  \end{center}
  \caption[]{Generic diagrams leading to the Coulomb singularity, the large 
             threshold logarithms, and the high-energy plateau, respectively.}
  \label{fig:coulomb}
\end{figure}

At the partonic level at least three sources of large corrections can be 
identified. Indicating the velocity of the produced heavy particles in their 
centre-of-mass system by $\beta$, two of these sources reside in the region 
near threshold ($\beta \ll 1$), from which an important part of the 
contributions to the hadronic cross-sections originates. First of all, the
exchange of (long-range) Coulomb gluons between the slowly moving
massive particles in the final state (see first generic diagram of 
Fig.~\ref{fig:coulomb}) leads to a singular correction factor 
$\sim \pi\as/\beta$, which compensates the LO phase-space suppression factor 
$\beta$. It should be noted, however, that the finite lifetimes of the 
squarks/gluinos reduce (screen) this effect considerably.
Secondly, as a result of the strong 
energy dependence of the cross-sections near threshold, large ``soft'' 
corrections $\sim \log^i(\beta^2)$ ($i=1,2$) are observed in the initial-state 
gluon-radiation contribution (see second generic diagram of 
Fig.~\ref{fig:coulomb}). At high energies, finally, the NLO partonic 
cross-sections behave asymptotically as a constant, rather than scaling with 
$1/s$ like the LO cross-sections. This is caused by the presence of almost 
on-shell IR gluons in space-like propagators, as appearing in the 
contributions from hard-gluon/quark radiation (see third generic diagram of 
Fig.~\ref{fig:coulomb}). All these effects can be calculated analytically in 
NLO, providing powerful checks on the partonic results 
\cite{BHSZprod,BHSZreport}.

After convolution of the partonic cross-sections with the relevant parton
densities, the hadronic cross-sections are obtained. When discussing LO and NLO
results, we calculate all quantities [$\as(Q_R^2)$, the parton densities, and
the partonic cross-sections] in LO and NLO, respectively. Bearing this in mind
the hadronic results can be summarized as follows.

\begin{figure}[bt]
  \begin{center}
  \vspace*{-1.7cm}
  \hspace*{-0.5cm}
  \epsfig{file=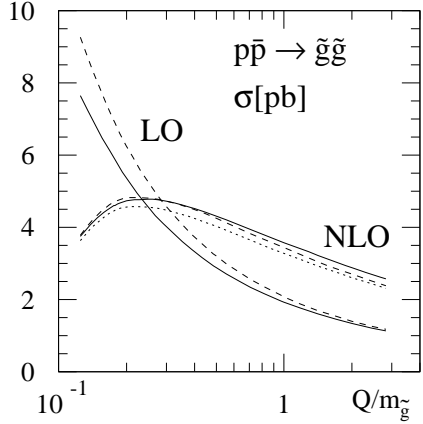,width=8cm}
  \vspace*{-2.3cm}
  \end{center}
  \caption[]{The scale and parton-density dependence of the LO and NLO 
             cross-sections for gluino-pair production at the Tevatron.
             Parton densities: GRV94 (solid), CTEQ3 (dashed), and MRSA'
             (dotted). Mass parameters: $\ms=280$~GeV, $\mg=200$~GeV, and 
             $m_t=175$~GeV.}
  \label{fig:sigscale1}
\end{figure}

{\it (i)} As is exemplified in Fig.~\ref{fig:sigscale1}, we find 
that the theoretical predictions for the production cross-sections are 
stabilized considerably by taking into account the NLO SUSY-QCD corrections. 
In all processes, for both the Tevatron and the LHC, it is observed that the 
dependence on the renormalization/factorization scale $Q$ ($Q=Q_R=Q_F$) is 
quite steep and monotonic in LO, whereas the $Q$ dependence is reduced by 
roughly a factor of 2.5--4 in NLO for reasonable variations of the scale. Even
a broad maximum develops for scales that are roughly a factor of 3--4 smaller 
than the average mass of the final-state particles. The variation of the 
cross-sections as a result of different NLO parametrizations of the parton 
densities is $\lsim 10\%$ at the Tevatron and 
$\lsim 13\%$ at the LHC, where the gluon densities are more important.
It should, however, be noted that considering such a sample of different parton
densities could lead to an underestimation of the actual
theoretical uncertainties. 
 
\begin{figure}[htb]
  \begin{center}
  \vspace*{-1.7cm}
  \hspace*{-0.5cm}
  \epsfig{file=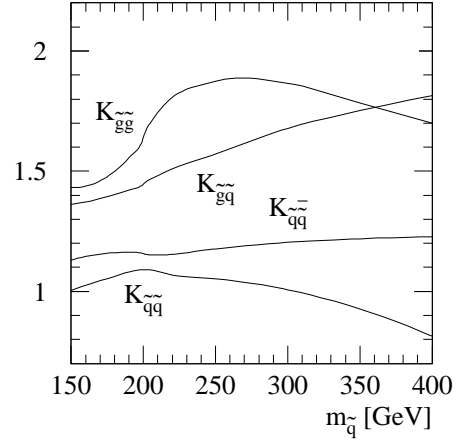,width=8cm}
  \vspace*{-2.0cm}
  \end{center}
  \caption[]{The K-factors for the Tevatron. Parton densities: GRV94 with 
             $Q$\,=\,average mass. Mass parameters: $\mg=200$~GeV and 
             $m_t=175$~GeV.}
  \label{fig:tevkfac}
\end{figure}

{\it (ii)} From now on we use GRV94 parton densities and conservatively take 
as default scale $Q$ the average mass of the produced particles. The 
K-factors, $\mbox{K}=\sigma_{NLO}/\sigma_{LO}$ , depend strongly on the 
process. This is exemplified in Fig.~\ref{fig:tevkfac} for the Tevatron.
For both collider types the NLO corrections are positive and large (up to 
$+90\%$) for the dominant production cross-sections, involving at least one 
gluino in the final state. The corresponding K-factors also exhibit a sizeable
dependence on the squark and gluino masses (especially the $\mg$ dependence of 
$\mbox{K}_{\gl\gl}$ is large). The NLO corrections for squark final states are 
moderate ($\lsim +30\%$). In view of the direct link between the total
cross-sections and the experimental determination of the squark/gluino masses 
in case of discovery, the inclusion of the NLO SUSY-QCD corrections is 
indispensable. 
 
{\it (iii)} Comparison of the NLO cross-sections with the cross-sections used 
in the Tevatron analysis (LO, EHLQ parton densities, and a scale $Q$ 
that equals the partonic centre-of-mass energy \cite{cdf1,d0}) reveals that
the NLO corrections raise the lower mass bounds for squarks and gluinos by 
$+10$~GeV to $+30$~GeV. 

{\it (iv)} Apart from the total cross-sections, also distributions with respect
to the rapidity ($y$) and transverse momentum ($p_t$) of one of the 
outgoing massive particles can be studied. The K-factors for these 
distributions are independent of $y$ for all practical purposes and hardly 
depend on $p_t$, except for large values of $p_t$ where the NLO corrections 
make the distributions somewhat softer. Consequently, multiplication of the 
LO distributions with the above-defined K-factors for the total cross-sections
approximates the full NLO results quite well.

\end{document}